\documentclass{elsart}

\usepackage{graphics}

\usepackage{graphicx}

\usepackage{epsfig}

\usepackage{amssymb}

\begin{document}

\begin{frontmatter}

\title{An Overview of the Basic Physical Properties of MgB$_2$}

\author{P.C. Canfield\corauthref{cor1},}
\corauth[cor1]{Corresponding author.}
\ead{canfield@ameslab.gov}
\author{S.L. Bud'ko, D.K. Finnemore}
\address{Ames Laboratory and Department of Physics and Astronomy, Iowa
State University, Ames, IA 50011, USA}

\begin{abstract}
The basic physical properties of MgB$_2$ have been well established over 
the past two years of intensive research.  At this point there is a general 
consensus about the values for the isotope shift, critical fields, most of 
the salient length scales, and general anisotropies.  In this paper we will 
review the determination of these parameters and set the stage for further, 
more detailed discussions of specific aspects of the physics of MgB$_2$.
\end{abstract}


\end{frontmatter}

\section{Introduction}

Given that the existence of superconductivity in MgB$_2$ was only announced in 
January of 2001, \cite{1} there has been an explosive growth in humanity's understanding 
of the basic physical properties of this simple binary compound over the past two 
years.  At this point in time there is a consensus about most of the fundamental 
properties of this material and a rapidly growing understanding of, as well as 
appreciation for, the dramatic differences between this extreme example of BCS 
superconductivity and other, older examples of intermetallic superconductivity.

In this paper we will summarize the status of our experimental understanding of 
the basic properties associated with the normal and superconducting states of MgB$_2$.

\section{Experimental procedures}

All of the data presented in this paper comes from measurements of polycrystalline 
samples of MgB$_2$.  Two basic types of samples were used:  sintered pellets and 
wire segments.  Both types of samples were synthesized by exposing boron to Mg vapor 
at, or near, 950$^{\circ}$C.  The sintered pellets were made by combining stoichiometric 
amounts of high purity magnesium and isotopically pure boron and reacting them in a 
sealed Ta tube.  \cite{2,3}  The wire segments were made by using a different initial form 
of boron:  boron filament.  In this case the boron filament was exposed to magnesium 
vapor by sealing it in a Ta tube with excess Mg and heating to or near 950$^{\circ}$C. 
\cite{4} This same method has been used to transform thin films of boron into MgB$_2$ 
thin films. \cite{5}  As those skilled in the art will realize, the steps outlined 
above provide a form preserving method of turning boron objects (filaments, films, tapes) 
into MgB$_2$ objects with similar morphologies. \cite{6}  Polycrystalline MgB$_2$ 
synthesized in this manner is single-phased and, as will be discussed below, 
manifests the properties of a high purity sample.  Further details about sample 
synthesis can be found in references \cite{2,3,4,5,6} as well as the paper 
by Ribeiro {\it et al.} in this volume.
	
Virtually all the measurements presented in this paper were performed on 
Quantum Design MPSM-5, MPMS-7, and PPMS-9 machines.  Electrical contact was made to the 
samples by using Epotek H20E silver epoxy and Pt wires.  High field magneto-transport 
data was measured at the National High Magnetic Field Laboratory in Los Alamos for 
magnetic field up to 18 T. \cite{7}

\section{Isotope effect measurements}

Immediately after the announcement of the discovery of superconductivity in MgB$_2$ 
the first question that had to be addressed was one of mechanism.  Whereas most 
researchers experienced with intermetallic compounds assumed that MgB$_2$ was the 
long sought after, extreme example of phonon-mediated, BCS superconductivity, some 
researchers, steeped in the lore of high-$T_c$ oxides, assumed that a 40 K superconductor 
had to stem from a more exotic coupling mechanism.  The simplest way to shed light on 
this question was to measure the boron isotope shift of $T_c$.  Looking at boron 
rather than magnesium was motivated by physical, economic and time considerations.  
Given the high $T_c$ of $\sim$40 K it seemed likely that the lightest (as well as most prevalent) 
element would be responsible for the phonons associated with the pairing of electrons.  
In addition, isotopically pure boron is relatively cheap given its natural abundance 
($\sim$80$\%$ $^{11}$B and 20$\%$ $^{10}$B) and the fact that its light atomic weight allow for 
separation via diffusion rather than requiring mass spectroscopy.  This fact means 
that many materials preparation labs, particularly ones working with neutron scatters 
(who require the use on non-absorbing $^{11}$B), have supplies of isotopically pure boron 
on hand.

\begin{figure}
\begin{center}
\includegraphics*[width=12cm]{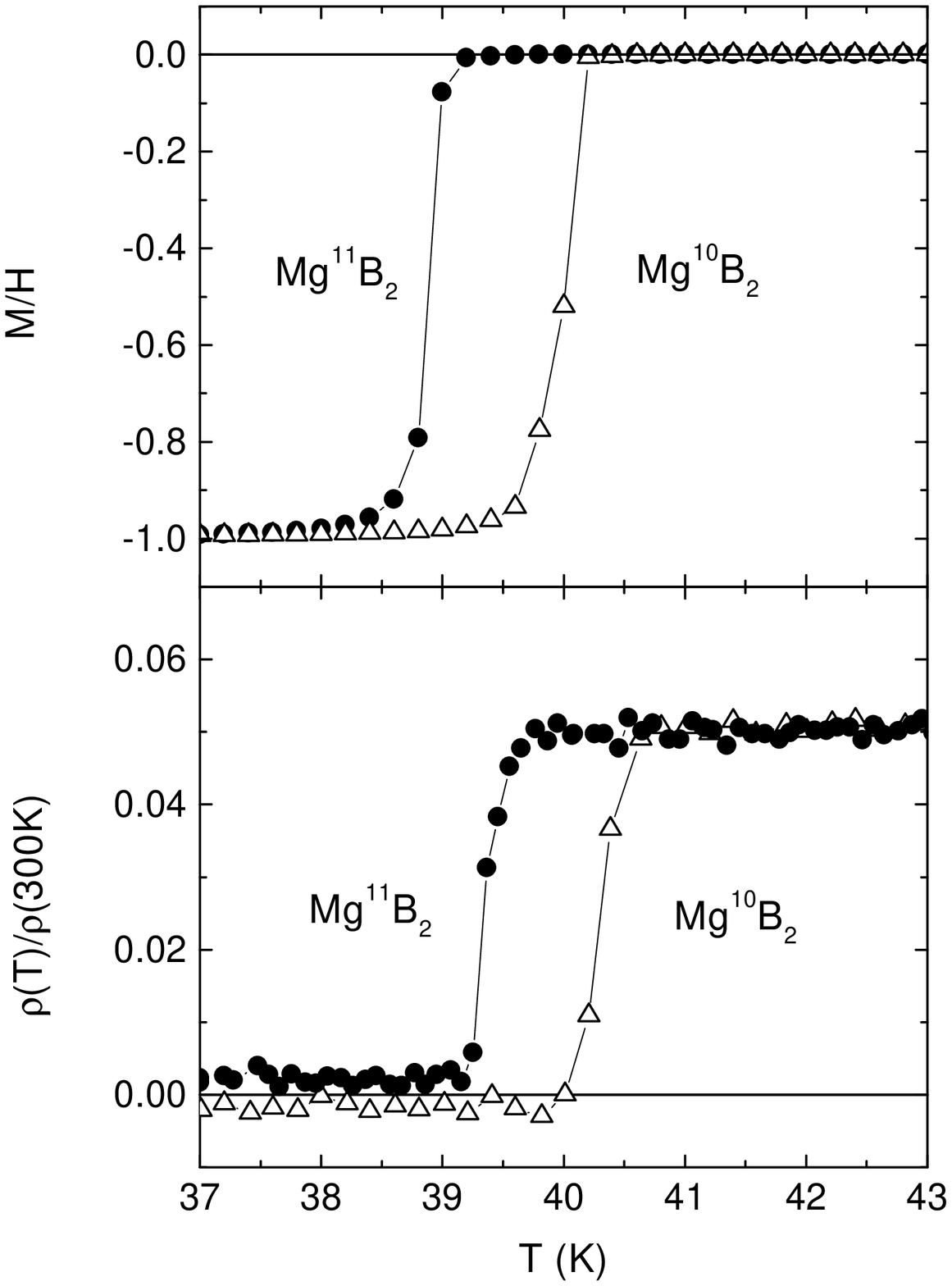}
\end{center}
\caption{Isotope shift as seen in: (upper panel) magnetization and (lower panel) resistance.}
\label{F1}
\end{figure}

Figure \ref{F1} presents temperature dependent electrical resistance and magnetization data 
taken on Mg$^{11}$B$_2$ and Mg$^{10}$B$_2$ samples.  A clear 1.0 K shift in $T_c$ can 
be seen in both measurements.  Specific heat data (not shown) taken on the same samples 
show the same 1.0 K shift.\cite{2}  This 1.0 K shift is to be compared with the simple 
prediction of the BCS model, $\Delta T_c$ $\sim$ 0.85 K, based upon 
$T_c \propto$ 1/(formula unit mass)$^{-1/2}$.  The partial isotope exponent for 
boron is $\alpha_B$ = 0.26. \cite{2}.  These data by themselves are not proof 
that Cooper pair formation in MgB$_2$ is mediated by phonons, but it is consistent with 
it and strongly reinforced the idea that MgB$_2$ was an extreme example of 
phonon-mediated BCS superconductivity.

Subsequent measurements \cite{8} confirmed the boron isotope shift and found that 
there was virtually no magnesium isotope shift.  All of the isotope shift data 
have subsequently explained within the context of phonon-mediated, BCS 
superconductivity. \cite{9,10,11}.

\section{Basic properties}

\begin{figure}
\begin{center}
\includegraphics*[width=12cm]{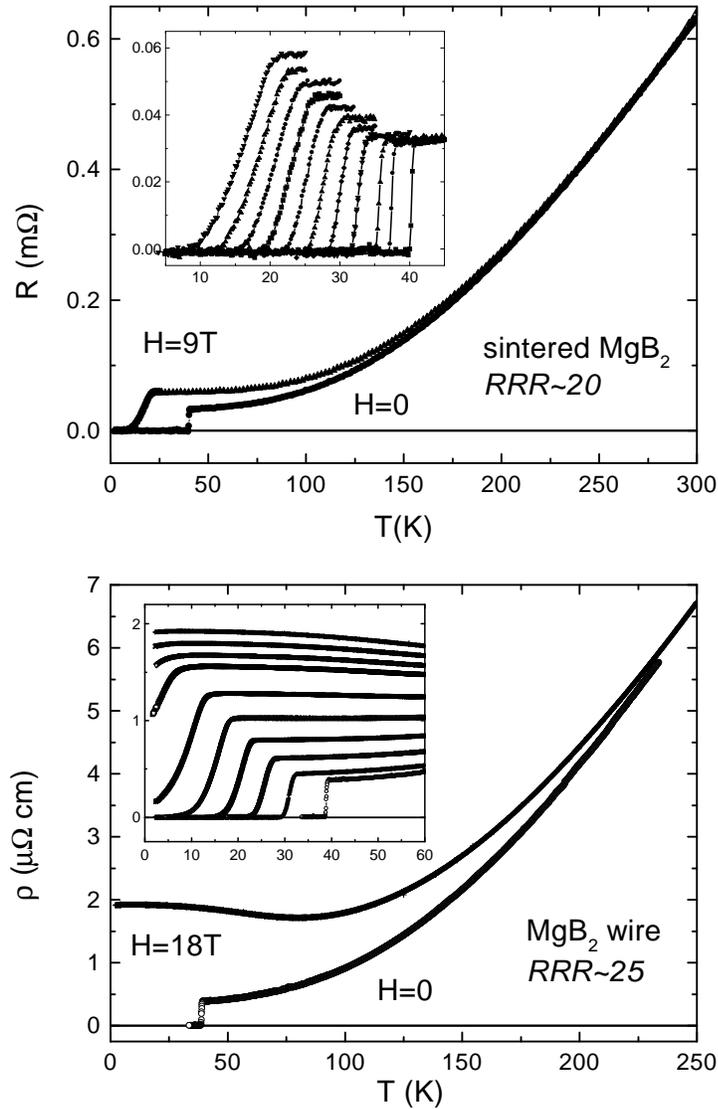}
\end{center}
\caption{Temperature dependence of electrical transport of MgB$_2$ at given applied 
magnetic fields:  (upper panel) resistance of sintered pellet in zero and 9 T applied 
field with inset showing low temperature data for applied fields of 0, 0.5, 1, 2, 
3, 4, 5, 6, 7, 8, and 9 T and (lower panel) resistivity of a dense wire segment in zero 
and 18 T applied field with inset showing low temperature data for applied fields of 
0, 2.5, 5, 7.5, 10, 12.5, 15, 16, 17, and 18 T.}
\label{F2}
\end{figure}

Once the question of mechanism has been addressed, the next set of questions that arise 
concern the basic properties associated with both the normal and superconducting state.  
Let's start with electrical transport.  Figure \ref{F2} (upper panel) presents the temperature 
dependent resistance of a piece of sintered pellet for various values of applied 
magnetic field.  Given that the density of the sintered pellet is not close to 100$\%$ 
we can only estimate the resistivity of MgB$_2$ at 42 K to be $\sim$ 1 $\mu\Omega$-cm based 
on these data.  On the other hand what can be clearly seen is that there is a substantial 
loose of scattering associated with cooling the sample from 300 K down to 40 K.  
In fact the residual resistivity ratio ($RRR$), $R$(300 K)/$R$(42 K), is close to 20. 
A consequence of this very low, normal state resistance is the substantial 
magneto-resistance observed in the normal state.  These effects can be seen even 
more clearly in the temperature dependent resistivity data from the MgB$_2$ wire 
segment (Fig. \ref{F2}, lower panel).  Given a sample density of better than 95$\%$ combined 
with a well-defined geometry the resistivity can be easily determined:  
e.g. $\rho$(300 K) = 9.6 $\mu\Omega$-cm and $\rho$(40 K) = 0.38 $\mu\Omega$-cm.  The 
temperature dependence of the wire sample is quantitatively similar to that seen 
for the sintered pellet and it manifests the same, large, normal-state 
magneto-resistivity.  

The value of the low temperature, normal-state resistivity is important for two 
rather different reasons.  If we take $v_F$ = 4.8$\times10^7$ cm/s we can extract 
an electronic mean free path of $l$ $\sim$ 600 $\AA$. \cite{4}  As will be shown below, 
when this length scale is compared to the superconducting coherence length, $\xi_0$, 
it becomes clear that MgB$_2$ is deep in the clean limit of superconductivity, 
i.e. $l \gg \xi_0$.  A low resistivity is also very important for possible 
applications, specifically high current ones.  An intrinsically small, 
low-temperature, normal state resistivity means that there is far less chance 
of quenching magnets made from this material.

\begin{figure}
\begin{center}
\includegraphics*[width=12cm]{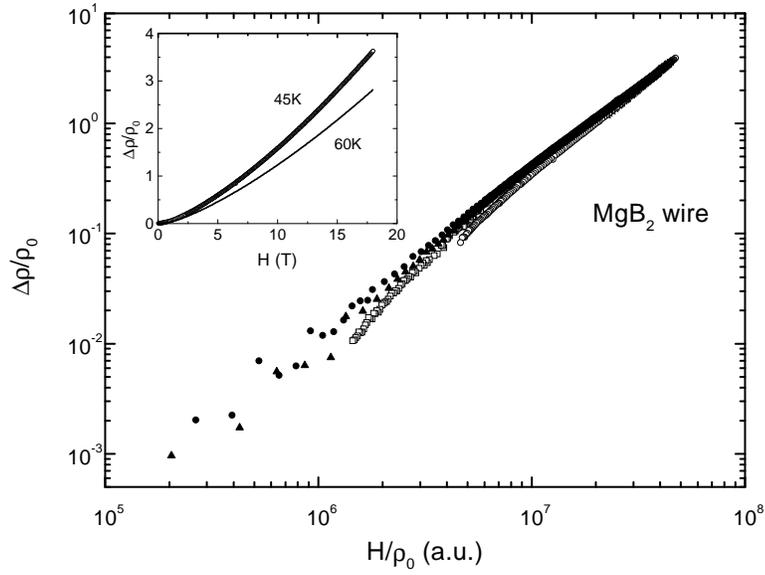}
\end{center}
\caption{Kohler's plot for MgB$_2$ wire sample:  open symbols from temperature 
dependent resistivity; filled symbols from field dependent resistivity taken at 
45 and 60 K as shown in the inset.  $\rho_0$ is the zero field resistivity, 
$\rho(T,0)$, and $\Delta\rho = \rho(T,H) - \rho(T, 0)$. 
}
\label{F3}
\end{figure}

Figure \ref{F3} presents the normal state magneto-transport data, shown in part in the 
lower panel of Fig. \ref{F2} as well as the inset of Fig. \ref{F3}, in the form of a Kohler 
plot. \cite{7,12}  The fact that all of these data fall onto a single manifold 
is consistent with MgB$_2$ being a simple ($s-$ and $p-$band) metal with a single 
salient scattering time in the normal state.  These data, along with the 
isotope effect data, help demonstrate that MgB$_2$ is indeed a representative 
example of intermetallic superconductivity, i.e. there is no need for more 
exotic, high-$T_c$-like, theoretical explainations.

It should be noted though, that whereas we have had no difficulty in preparing 
high-purity, low resistivity, polycrystalline samples of MgB$_2$, other groups had 
difficulty in reproducing these results.  Residual resistivity values as high as 
70 $\mu\Omega$-cm \cite{1} or 20 $\mu\Omega$-cm \cite{13} have been reported.  
In addition it was suggested that metallic magnesium in parallel with the 
MgB$_2$ could be responsible for the low resistivities we have 
found. \cite{14,15}  This controversy spurred us to examine the possible sources 
of this apparent variation in sample quality (see reference \cite{3} as well as 
paper by R. Ribeiro {\it et al.} in this volume).  We found that the use of very 
high purity boron is vital to the formation of low resistivity samples.  We also 
found that for the stiochiometric pellets as well as the wire segments formed via 
exposure to boron filament to Mg vapor the low resistivity is not associated 
with free magnesium, but rather is an intrinsic feature of high purity MgB$_2$.  
Fortunately as more and more data on MgB$_2$ (polycrystals, films, and single 
crystals) becomes available there values of both room temperature and residual 
resistivity are asymptotically approaching the values presented in 
Fig. \ref{F2}. \cite{16,17,18}

\begin{figure}
\begin{center}
\includegraphics*[width=12cm]{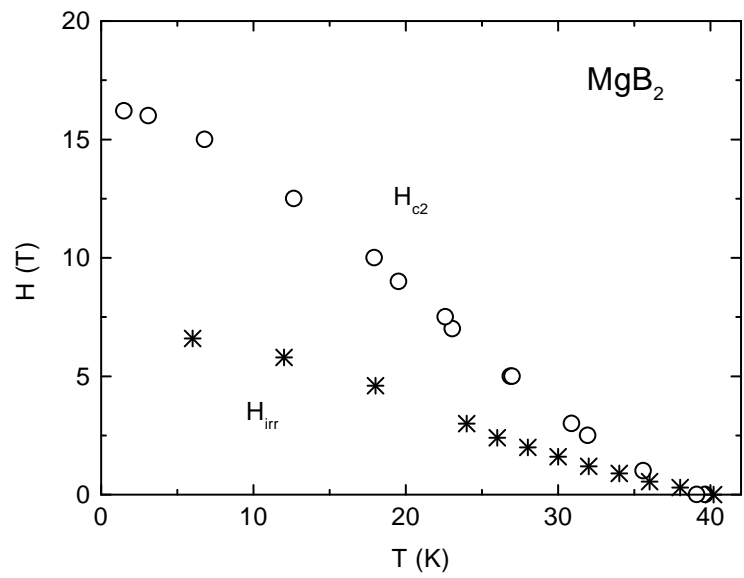}
\end{center}
\caption{Temperature dependence of $H_{c2}$ (open circles) and $H_{irr}$ (stars).}
\label{F4}
\end{figure}

The magneto-transport data shown in Fig. \ref{F2} can also be used to extract information 
about the superconducting state, specifically the upper critical field, $H_{c2}(T)$.  
The $H_{c2}(T)$ data are plotted in Fig. \ref{F4}.  $H_{c2}(T)$ rises to slightly above 16 T 
at 1.5 K. \cite{7}  From this value we can extract an estimate of $\xi_0$ of $\sim$ 50 $\AA$, 
a value that is indeed much less than the electronic mean free path, $l$ $\sim$ 600 $\AA$.  
As mentioned above, given that $l \gg \xi_0$, MgB$_2$ is deep within the clean limit 
of superconductivity.

\begin{figure}
\begin{center}
\includegraphics*[width=12cm]{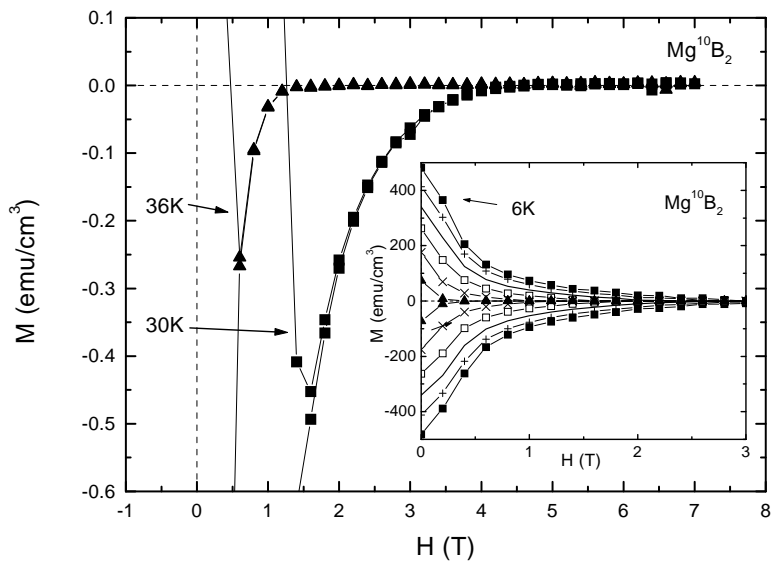}
\end{center}
\caption{Expanded view of magnetization vs. field to show the reversible range 
as well as $H_{irr}$.  The inset shows the full range of magnetization data up 
to 3 T at temperature intervals of 6 K.}
\label{F5}
\end{figure}

There are several other field scales of interest:  the irreversibility field, 
$H_{irr}$ and the thermodynamic critical field, $H_c$.  Figure \ref{F5} presents $M(H)$ 
data that clearly show the irreversibility field for representative temperatures.  
The $H_{irr}(T)$ curve deduced from these data is presented as the lower curve 
in Fig. \ref{F4}.  As can be seen, $H_{irr}$ is approximately one half of $H_{c2}$.  
Whereas the values of $H_{c2}$(1.5 K) $\sim$ 16 T and $H_{irr}$(5 K) $\sim$ 7 T are neither 
very low nor very high, there are already examples of dirtier samples for which the 
low temperature $H_{c2}$ $\sim$ 30 T and the low temperature $H_{irr}$ $\sim$ 15 T. 
\cite{19,20}  One of the current, key areas of applied research on MgB$_2$ is 
focusing on the attempt to understand how pinning, as well as $\kappa$, in this 
material can be simply and reproducibly controlled.

\begin{figure}
\begin{center}
\includegraphics*[width=12cm]{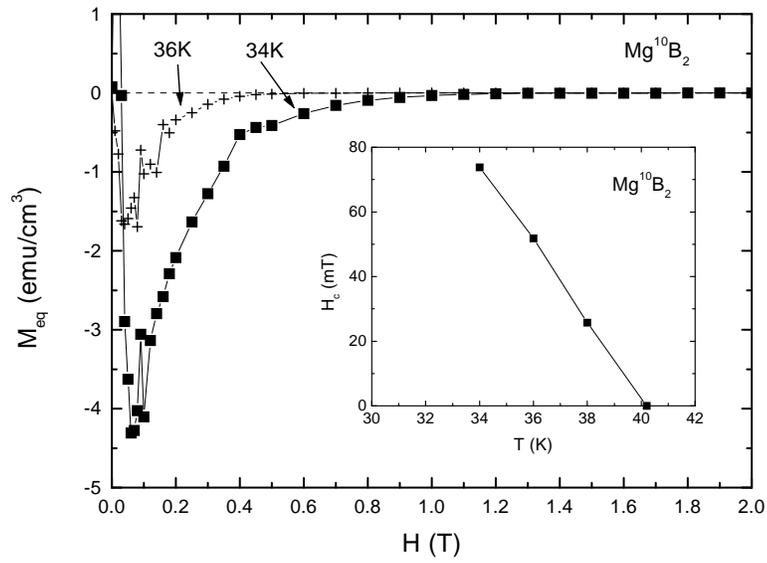}
\end{center}
\caption{Equilibrium magnetization as a function of applied field (as described in text).  
Inset:  Thermodynamic critical field $H_c(T)$ for $T$ close to $T_c$.}
\label{F6}
\end{figure}

The thermodynamic critical field can be inferred, at least for $T$ near $T_c$, 
from magnetization loops. \cite{12}  The inset to Fig. \ref{F6} presents $H_c(T)$ for 
$T$ close to $T_c$.  $H_c(T)$ was determined by integrating the area under 
the equilibrium magnetization versus field curve where 
$M_{eq} = (M_{inc} + M_{dec})/2$, and $M_{inc}$ and $M_{dec}$ are the increasing 
field magnetization and decreasing field magnetization respectively. \cite{12} 
The $H_c(T)$ data can be used to estimate another length scale, the London 
penetration depth, $\lambda$, by using the simple relations 
$\kappa = H_{c2}/[1.41 H_c] = \lambda/\xi_0$.  Using the $H_{c2}(T)$ data from 
Fig. \ref{F4} as well as the $H_c(T)$ data from Fig. \ref{F6} we find that $\kappa$ $\sim$ 25 and 
$\lambda$ $\sim$ 1400 $\AA$.

\begin{figure}
\begin{center}
\includegraphics*[width=12cm]{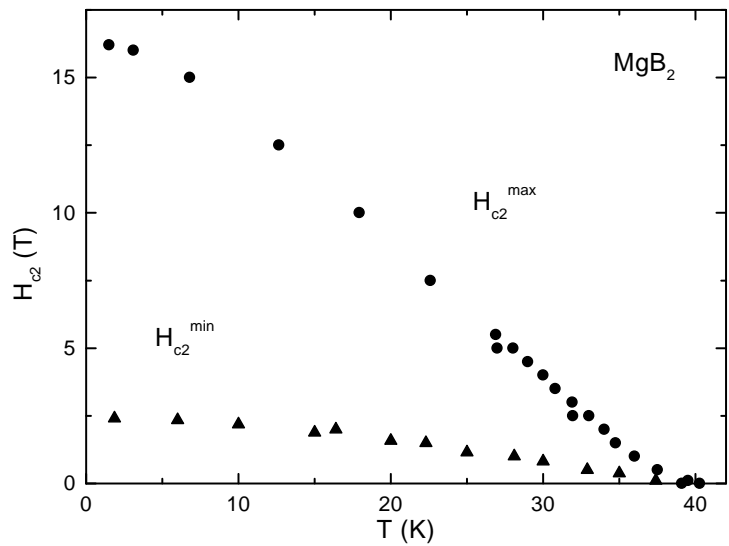}
\end{center}
\caption{Anisotropic $H_{c2}(T)$ inferred from measurements on polycrystalline samples.  
(As described in references \cite{23} and \cite{24}.)  $H_{c2}^{min}$ is the upper critical 
field for $H$ parallel to the crystallographic $c$-axis and $H_{c2}^{max}$ is the upper 
critical field for $H$ perpendicular to the $c$-axis.}
\label{F7}
\end{figure}

The values of $l$, $\xi_0$, $\lambda$ and $\kappa$ derived above are average 
values.  There has been no effort made to account for the anisotropy associated 
with the hexagonal unit cell of MgB$_2$.  This means that these values should be 
taken as bench marks rather than definitive, final values.  The reason for this 
caveat is the fact that there are very large anisotropies in many of the properties 
of MgB$_2$.  Perhaps the most conspicuous is the anisotropy in the value of 
$H_{c2}(T)$. \cite{21,22,23,24}  Whereas early estimates of this anisotropy, 
$\gamma \equiv H_{c2}^{basal~plane}/H_{c2}^{c-axis}$, were rather low 
($\gamma$ = 1.73), \cite{21} measurements on unoriented, high purity powders 
allowed us to determine the anisotropic $H_{c2}(T)$ in detail. \cite{23,24}  
Figure \ref{F7} presents the temperature dependence of the anisotropic $H_{c2}(T)$ curves.  
The lower curve is for the crystallites with the field aligned along the 
crystallographic $c$-axis and the upper curve is for the crystallites with the 
field aligned along the basal plane. \cite{21,22}  At low temperatures 
$\gamma$ approaches a remarkably high value between 6 and 7. Over the past several 
months, detailed measurements on small single crystalline samples have confirmed 
not only the size of the low temperature anisotropy, but also the full temperature 
dependence of $\gamma(T)$. \cite{25,26}  The low temperature value of the 
anisotropy, $\gamma$ = 6-7, is in excellent agreement with the anisotropic 
$H_{c2}(T)$ that would be anticipated if only the two dimensional, cylindrical 
Fermi surface is responsible for the low-$T$, $H \ge$ 2 T, superconductivity.
\cite{23}  If this is the case, as recent experiments seem to confirm, 
\cite{27} then $H_{c2}^{max}/H_{c2}^{min} = (\langle v_{ab}^2\rangle/\langle v_c^2\rangle)^{1/2}$, 
assuming that the gap over the cylinder is essentially constant. Using the anisotropic 
Fermi velocities for the 2-D cylinder \cite{23} we find that this ratio to 
be $\sim$ $\surd$ 40 which is in excellent agreement with the measured value 
of $\gamma$ at low temperatures.  A fuller and more detailed presentation of the 
temperature dependence of anisotropies in MgB$_2$ will be presented by 
Kogan {\it et al.}, as well as others, later in this volume.

\section{Critical current density}

\begin{figure}
\begin{center}
\includegraphics*[width=12cm]{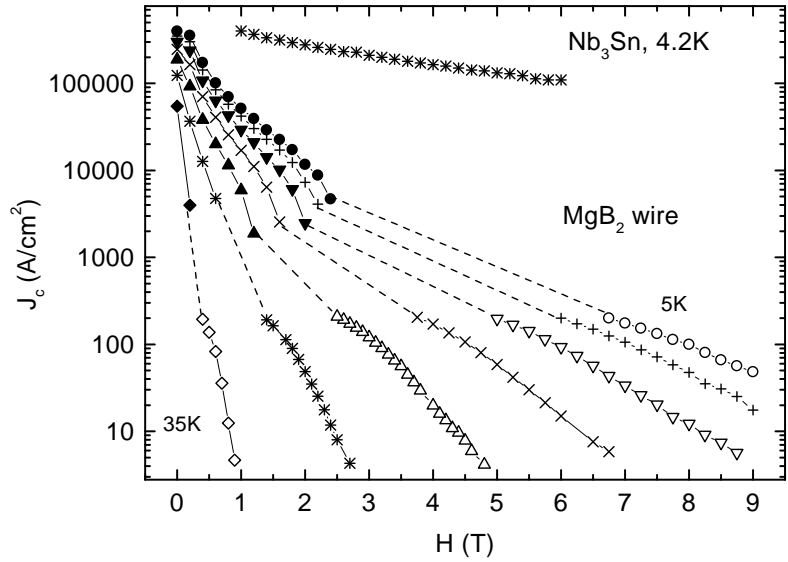}
\end{center}
\caption{Superconducting critical current density, $J_c$, as a function of applied 
field every 5 K in the 5 - 35 K range.  Lower current density values, open symbols, 
were directly determined via $V(I)$ measurements.  Higher current density values, 
filled symbols, were determined via a Bean-model analysis of $M(H)$ loops.  $J_c(H)$ 
data for Nb$_3$Sn taken at 4.2 K are shown for comparison.}
\label{F8}
\end{figure}

The $M(H)$ isotherms that were used to determine $H_{irr}(T)$ as well as $H_c(T)$ also 
yielded information about the critical current density, $J_c$.  Figure \ref{F8} presents 
$J_c(H)$ for a variety of representative temperatures.  The higher values of 
$J_c(H,T)$ were inferred from magnetization data using the Bean model \cite{4,12,28} 
and the lower values of $J_c$ were directly measured via $V(I)$ measurements. 
\cite{4}  For comparison $J_c(H)$ data for Nb$_3$Sn are also shown for $T$ = 4.2 K. 
\cite{29} 

Several points are worth noting.  First of all, even though these $J_c$ values are 
lower than those for Nb$_3$Sn at 4-5 K, the $J_c$ values are not all that small, 
especially when you consider that the sample was high purity MgB$_2$.  In addition, 
at 20 K Nb$_3$Sn is in the normal state (i.e. $J_c$ = 0) whereas the $J_c$ values for 
MgB$_2$ are still sizable for low to intermediate fields.  Secondly, $J_c(H,T)$ is an 
extrinsic property that can be increased dramatically by judicious choices of impurities 
and defects to act as pinning sites.  Over the past year $J_c$ values in wire segments have 
increased by approximately an order of magnitude.  Recent work by Finnemore {\it et al.} 
(see later in this volume) on Ti and C doped wires has shown that the low temperature, 
low field value of $J_c$ can be increased to mid- 10$^6$ A/cm$^2$.  Comparable values 
have been found for thin film samples. \cite{19}.  Hopefully further increases in 
$J_c$, not only for low fields, but also for higher fields will continue over the next 
few years.

\section{Summary}

MgB$_2$ is a poster child for two very different and important observations about 
solid state physics.  First, the field of solid state physics has come a fantastically 
long way given that the basic physical properties of MgB$_2$, as well as our theoretical 
understanding of them, were well delineated within less than a year of its rediscovery.  
This is a remarkable fact and should not be given short shrift.  Secondly, superconductivity 
in this simple binary compound lay undiscovered until the new millennium, despite extensive 
searches over the past 80 years, and although superconductivity in this material is now well 
understood, it was not predicted {\it a priori}.  It is vital to experimentally search 
for (i) new materials and (ii) new properties / groundstates in old ones.  The continued 
search for new materials and new properties in binary, ternary, quaternary, and even more 
complex compounds is one of the most important engines driving this field of science.

We gratefully acknowledge useful interactions with A. H. Lacerda, M-H. Jung, N. Anderson, 
G. Lapertot, R. Ribiero, C. Petrovic, J. E. Ostenson, V. G. Kogan and V. P. Antropov.  
We also thank the U.S. D.O.E. Office of Basic Energy Sciences for the flexibility in funding 
that to allowed us to chase after this wonder material. Ames Laboratory is operated for 
the U.S.
Department of Energy by Iowa State University under contract No.
W-7405-ENG-82.  The work was supported by the Director of Energy Research,
Office of Basic Energy  Sciences, U. S. Department of Energy.

\clearpage

\clearpage

\clearpage

\clearpage

\clearpage

\clearpage

\clearpage

\clearpage

\end{document}